\documentclass[aps,jcp,preprint,groupedaddress]{revtex4-1}
\usepackage{amsmath}
\usepackage{color}
\usepackage{wasysym}	%\apprle
\usepackage{setspace}

\usepackage{graphicx}% Include figure files
\usepackage{dcolumn}% Align table columns on decimal point
\usepackage{bm}% bold math
\usepackage{longtable}

\newcommand{\ea}{{\it et al.}}

\begin{document}
%\preprint{APS/123-QED}

\title{Molecular Simulation of Caloric Properties of Fluids Modelled by Force Fields with Intramolecular Contributions: Application to Heat Capacities}

\author{William R. Smith}
\affiliation{Department of Mathematics and Statistics, College of Physical and Engineering Science, University of Guelph, Guelph, ON, Canada N1G 2W1}
\affiliation{Faculty of Science, University of Ontario Institute of Technology, Oshawa, ON, Canada, L1H 7K4}
\email{bilsmith@uoguelph.ca}
\author{Jan Jirs\'{a}k}
\affiliation{Institute of Chemical Process Fundamentals, Academy of Sciences, 165 02 Prague 6, Czech Republic}
\affiliation{Department of Chemistry, Faculty of Science, J. E. Purkyn\v{e} University, 400 96  \'{U}st\'{\i} n. Lab., Czech Republic}
\author{Ivo Nezbeda}
\affiliation{Institute of Chemical Process Fundamentals, Academy of Sciences, 165 02 Prague 6, Czech Republic}
\affiliation{Department of Chemistry, Faculty of Science, J. E. Purkyn\v{e} University, 400 96  \'{U}st\'{\i} n. Lab., Czech Republic}
\author{Weikai Qi}
\affiliation{Department of Mathematics and Statistics, College of Physical and Engineering Science, University of Guelph, Guelph, ON, Canada N1G 2W1}


\begin{abstract}
The calculation of caloric properties such as heat capacity, Joule-Thomson coefficients and the speed of sound by classical force-field-based molecular simulation methodology has received scant attention in the literature, particularly for systems composed of complex molecules whose force fields (FFs) are characterized by a combination of intramolecular and intermolecular terms (referred to herein as ``flexible FFs").  The calculation of a thermodynamic property for a system whose molecules are described by such a FF involves the calculation of the residual property prior to its addition to the corresponding ideal-gas (IG) property, the latter of which is separately calculated, either using thermochemical compilations or nowadays accurate quantum mechanical calculations.  Although the simulation of a volumetric residual property proceeds by simply replacing the intermolecular FF in the rigid molecule case by the total (intramolecular plus intermolecular) FF, this is not the case for a caloric property.  We discuss the methodology required in performing such calculations, and focus on the example of the molar heat capacity at constant pressure, $c_P$, one of the most important caloric properties.  We also consider three approximations for the calculation procedure, and illustrate their consequences for the examples of the relatively simple molecule 2-propanol, ${\rm CH_3CH(OH)CH_3}$, and for monoethanolamine, ${\rm HO(CH_2)_2NH_2}$, an important fluid used in carbon capture.
\end{abstract}

%\clearpage

\maketitle

\section{Introduction}

Molecular simulations using classical Molecular Dynamics (MD) and Monte Carlo (MC) methodology are becoming increasingly employed to model both thermodynamic and transport properties in many  scientific and application areas~\cite{Ungerer2007,Maginn2010,Meunier2011,Biscay2011,Dror2012}.  In parallel with the rapid development of computer hardware over the past several decades, their thermodynamic capabilities have evolved from the calculation of simple volumetric properties for pure fluid systems composed of spherically symmetric molecules to additional and more sophisticated properties for multi-component mixtures composed of large and complex molecules with intramolecular degrees of freedom.   Caloric properties are an important class of such properties, whose members include $A,G,S,U,H,C_V,C_P, \mu_{\rm JT}, w_s$ ($A$ is the Helmholtz energy, $G$ is the Gibbs energy, $S$ is the entropy, $U$ is the internal energy, $H$ is the enthalpy, $C_V$ is the constant volume heat capacity, $C_P$ is the isobaric heat capacity, $\mu_{\rm JT}$ is the Joule-Thomson coefficient and $w_s$ is the speed of sound).

A primary goal of molecular simulations is material property prediction, both in cases for which experimental measurements for existing substances are unavailable or inaccurately known and for proposed new substances for which no experimental measurements exist.  The underlying methodologies and the molecular models utilized within such simulations must first be validated by means of comparisons with experimental data in situations for which they are available.  In the literature, the vast majority of such comparisons have involved volumetric $(PVT)$ properties ($P$ is the pressure, $V$ the volume and $T$ the absolute temperature) and phase equilibrium properties, simulation methodologies for which are now well established.  The former properties are directly available experimentally, and the latter involve experimentally measurable properties that (for pure substances) are consequences of the equality of $G, T$ and $P$ for the coexisting phases.  Much less attention has been focussed on predictions of directly measurable caloric properties, the most important of which are the heat capacities.  An indication of their fundamental importance is the fact that all thermodynamic first derivatives involving the quantities $\{A,G,S,U,H,P,V,T\}$, can be expressed in terms of a set of three such derivatives.  Bridgman~\cite{Bridgman1914} developed a set of compact equations to express any such first derivative in terms of the members of the ``fundamental set" $\{C_P, \alpha_P, \beta_T\}$, where $\alpha_P$ is the volumetric thermal expansion coefficient and $\kappa_T$ is the isothermal compressibility; $\alpha_P$ and $\kappa_T$ are both volumetric properties, and $C_P$ is the only caloric property in this set.

The basic input to a classical molecular simulation is a Force Field (FF), a mathematical model with specified parameter values describing the potential energy of the interacting  molecules as a sum of intermolecular and intramolecular contributions.  A small molecule may be modeled as a rigid geometrical body  and its FF contains no intramolecular terms; we henceforth call this a {\it rigid FF}, based on a {\it rigid FF model}.  A larger and more complex molecule requires a FF model that includes intramolecular contributions; we henceforth call this a {\it flexible FF model}, and the resulting FF a {\it flexible FF}.

Under the usual pairwise additivity approximation, the sum of the intramolecular and intermolecular configurational energy of a simulated system of $N$ molecules, $\cal{U}_{\rm total}({\bf z})$, is expressed as
\begin{eqnarray}
{\cal{U}}_{\rm total}({\bf z}) & = & {\cal U}_{\rm intra}({\bf z}) +{\cal U}_{\rm inter}({\bf z}) \label{eq:0}\\
                   & = &  \sum_{i=1}^{N} u_{{\rm intra}}({\bf s}_i) + \sum_{i=1}^{N-1}\sum_{j=i+1}^{N}u_{\rm inter}({\bf z}_i,{\bf z}_j)\label{eq:2}
\end{eqnarray}
where ${\bf z}$ denotes all molecular coordinates, ${\bf z}_i\equiv({\bf q}_i,{\bf s}_i)$ for $i=1\ldots N$,  ${\bf s}_i$ denotes the `internal', molecular frame coordinates of molecule $i$, and ${\bf q}_{i}\equiv({\bf r}_i,\boldsymbol{\omega}_i)$ denotes the `external', laboratory frame coordinates, describing the position, ${\bf r}_i$, and orientation, $\boldsymbol{\omega}_i$, of molecule $i$.

An example of a flexible FF model is the OPLS family, due originally to Jorgensen\cite{Jorgensen1986}, which is in widespread use.  The OPLS FF model contains four intramolecular contributions, and is expressed as:
\begin{equation}
u_{\rm intra}^{\rm OPLS}({\bf s}_i) = u_{\rm bond\,\, stretching} + u_{\rm bond\,\, bending} + u_{\rm bond\,\, torsion} + u_{\rm intra\; non-bonded} \label{eq:OPLS}
\end{equation}

Most of the relatively small number of simulation studies of heat capacities and their residual values, $C_P^{\rm res}$ and $C_V^{\rm res}$, have been performed for systems modelled by rigid FFs, {\it e.g.}, ~\cite{Kronome1999,Pineiro2008,Gomez2013}. Jorgensen was among the first to consider flexible FFs, when he calculated the enthalpy of vaporization, $\Delta_{\rm vap}H$, and $C_P$ values for several fluids.  He proposed using the former quantity as a training set property in fitting the intermolecular OPLS FF parameters to experimental data~\cite{Jorgensen1986,Jorgensen1996}.  This methodology has been widely adopted by subsequent workers in the development and application of OPLS FFs.  Other calculations of heat capacities for flexible FFs  include calculations of $C_V$ for alkanes up to heptane by Escobedo and Chen~\cite{Escobedo2001},  of $C_P^{\rm res}$ and $C_P$ for the refrigerant HFO-1234yf by Raabe and Maginn~\cite{Raabe2010a}, of $C_P$ for a range of pure fluids by Caleman \ea\cite{Caleman2012}, and of $C_P$ and Joule-Thomson coefficients of fluids important in the oil and natural gas processing industry by Ungerer and co-workers~\cite{Lagache2001,Lagache2004,Ungerer2005,Ungerer2007,Biscay2011,Ungerer2015}.  Careful reading of these works reveals that they used various approximations for the heat capacity calculations, which we will indicate in Section \ref{sec:2.3}.

In spite of the widespread use and importance of flexible FFs for modeling the behaviour of fluids, the general methodology for calculating their residual thermodynamic properties from simulations has not been described in the literature.  The goal of this paper is to describe this methodology, which also permits the calculation of the total thermodynamic property values by adding their separately obtained ideal-gas values.  We emphasize caloric properties and consider a pure fluid, but the results are readily extended to mixtures. We will show that residual volumetric properties may be correctly calculated by treating the sum of the intramolecular and intermolecular potentials in the same way that the intermolecular potential is treated in the case of a rigid FF, and we will show that caloric residual properties must be calculated according to a different procedure.   We also consider three  approximations to the methodology.  Illustrative examples are provided by calculations of the caloric properties $H$ and $C_P$ of 2-propanol: ${\rm CH_3CH(OH)CH_3}$ and of monoethanolamine (MEA): ${\rm HO(CH_2)_2NH_2}$.  $C_P$ and the enthalpy of vaporization, $\Delta_{\rm vap}H$, were first considered by  Jorgensen~\cite{Jorgensen1986,Jorgensen1996}, and MEA is an important solvent in carbon capture processes~\cite{Tan2016c}, and illustrative of a fluid consisting of a larger and more complex molecule.

In the next section of the paper, we derive the methodology, and in the following section we describe the systems considered and their molecular models, in addition to the technical details of the simulations.  The subsequent section describes our results and the final section gives our conclusions.

\section{Thermodynamic Property Calculations}

\subsection{Rigid FFs}
We first briefly review the simulation methodology for the calculation of total and residual properties in the case of a rigid FF, using a development that extends naturally to that for a flexible FF. For simplicity, we consider a pure fluid system described by the canonical ensemble at specified $(N,V,T)$, where $N$ is the number of particles.

In the case of a rigid FF, the intramolecular and intermolecular degrees of freedom of the molecules are assumed  not to mutually interact.  The IG partition function may thereby be factored out, and the total and IG partition functions expressed as ({\it e.g.}, MacQuarrie~\cite{McQuarrie2000})
\begin{eqnarray}
Q_{\rm total} & = & Q_{\rm IG}Q_{\rm config} \label{eq:rigidFF1}\\
Q_{\rm IG}  & =  & Q_{\rm trans}Q_{\rm frot}Q_{\rm irv}Q_{\rm electronic}Q_{\rm nuclear} \label{eq:rigidFF33}
\end{eqnarray}
where
$Q_{\rm trans}$ is the translational partition function; $Q_{\rm frot}$ is the partition function corresponding to rotation of the entire molecule; $Q_{\rm irv}$ is the combined internal rotational-vibrational partition function; $Q_{\rm electronic}$ is the electronic partition function;
$Q_{\rm nuclear}$ is the nuclear partition function; and $Q_{\rm config}$ is the configurational partition function, determined by the intermolecular FF $u_{\rm inter}({\bf z})$.

The translational and configurational components of $Q_{\rm total}$ are given by
\begin{eqnarray}
Q_{\rm trans} & = & \frac{1}{N!}V^N\left[\frac{2\pi m}{\beta h^2}\right]^{3N/2} \label{eq:trans}\\
Q_{\rm config} & = & \int \exp(-[\beta \cal{U}_{\rm inter}({\bf z})]){\bf dz} \label{eq:rigidQ}
\end{eqnarray}
where the symbolic differential ${\bf dz}$ in Eq. (\ref{eq:rigidQ}) is assumed to include the appropriate Jacobian and a normalization factor to yield $\int{\bf dz} = 1$.
$m$ is the molecular mass, $\beta = 1/(k_BT)$ where $k_B$ is Boltzmann's constant and $h$ is Planck's constant.   For $Q_{\rm frot}$, the three possible different cases are described in standard textbooks~\cite{McQuarrie2000}.

In what follows, we focus on $Q_{\rm irv}$, and write Eqs. (\ref{eq:rigidFF1})  and (\ref{eq:rigidFF33}) as
\begin{eqnarray}
Q_{\rm total} & = & Q^0_{\rm IG}Q_{\rm irv}Q_{\rm config} \label{eq:rigidFF35}\\
Q_{\rm IG}  & =  & Q^0_{\rm IG}Q_{\rm irv} \label{eq:rigidFF34}
\end{eqnarray}
Eq. (\ref{eq:rigidFF34}) formally defines the quantity $Q^0_{\rm IG}$.

The total value of the Helmholtz energy, $A$, is related to $Q_{\rm total}$ and its component terms in Eq. (\ref{eq:rigidFF1}) by
\begin{eqnarray}
\beta A  & = &  -\ln Q_{\rm IG} - \ln Q_{\rm config}  \label{eq:10}\\
 & = & \beta A^{\rm IG}(\rho,T) + \beta A^{\rm res}(\rho,T) \\
 & = &  N \beta \tilde{a}^{\rm IG}(\rho,T) + \beta A^{\rm res}(\rho,T) \label{eq:Aexact}
\end{eqnarray}
where $\rho=N/V$ is the density, $\tilde{a}_{\rm IG}(T;\rho)$ is the value of $A_{\rm IG}$ per molecule (separately available or obtained from a quantum mechanical calculation) and superscript res denotes a value in excess of that of the ideal gas at the given $\rho$ and $T$.

The important point for our purposes is that $\beta A^{\rm res}$ arises from ${\cal U}_{\rm inter}$ and hence from $u_{\rm inter}$.  In the case of a rigid FF, the pressure and the internal energy, $U$, given by
\begin{eqnarray}
\beta P  =  -\left(\frac{\partial \beta A}{\partial V}\right)_{N,T} & = & \frac{N}{V}
 - \left(\frac{\partial \beta A_{\rm config}}{\partial V}\right)_{N,T} \label{eq:pressure0}\\
& \equiv  & (\beta P)^{\rm IG} +  (\beta P)^{\rm res}  \label{eq:pressure} \\
U = \left(\frac{\partial \beta A}{\partial \beta}\right)_{N,V} & = & \left(\frac{\partial \beta A_{\rm IG}}{\partial \beta}\right)_{N,V} + \left(\frac{\partial \beta A_{\rm config}}{\partial \beta}\right)_{N,V}\\
& = & N\beta \tilde{u}^{\rm IG}(T) + U^{\rm res}(\rho,T) \label{eq:Uexact}
\end{eqnarray}

\subsection{Flexible FFs}
For more complex molecules (and even for small molecules at extreme conditions), the internal and intermolecular degrees of freedom may no longer be treated as completely separable, which is taken into account by modeling the quantity $Q_{\rm irv}$ in Eqs. (\ref{eq:rigidFF35}) and (\ref{eq:rigidFF34}) by an intramolecular potential energy term $\cal{U}_{\rm intra}({\bf z})$, and those equations become
\begin{eqnarray}
Q_{\rm total}  & =  & Q^{0}_{\rm IG}Q^{\rm total}_{\rm config} \label{eq:rigidFF7}\\
Q_{\rm IG} & = & Q_{\rm IG}^{0} Q^{\rm total,IG}_{\rm config} = Q_{\rm IG}^{0} Q^{\rm intra,IG}_{\rm config} \label{eq:QID2}
\end{eqnarray}
where
\begin{equation}
Q^{\rm total}_{\rm config}  =  \int \exp(-[\beta \cal{U}_{\rm total}({\bf z})]){\bf dz} \label{eq:FlexQ}
\end{equation}
Similarly to the case of Eq. (\ref{eq:rigidFF34}), a rigid FF, Eq. (\ref{eq:QID2}) provides a formal definition of the quantity $Q^0_{\rm IG}$.

Dividing Eq. (\ref{eq:rigidFF7}) by Eq. (\ref{eq:QID2}) to eliminate $Q^{0}_{\rm IG}$ yields
\begin{equation}
Q_{\rm total} = \left(\frac{Q_{\rm IG}}{Q^{\rm intra,IG}_{\rm config}}\right)Q^{\rm total}_{\rm config} \label{eq:finalQQ}
\end{equation}
Eq. (\ref{eq:finalQQ}) then gives the counterparts of Eqs. (\ref{eq:10})-(\ref{eq:Aexact}):
\begin{eqnarray}
\beta A  \equiv -\ln Q_{\rm total} & = & -\ln Q_{\rm IG} - \ln Q^{\rm total}_{\rm config} + \ln Q^{\rm intra,IG}_{\rm config}  \label{eq:first} \\
& = & \beta A^{\rm IG}(\rho,T)   + \beta A^{\rm total}_{\rm config}(\rho,T)  -N\beta \tilde{a}_{\rm intra}^{\rm IG}(T) \label{eq:finalnew0} \\
& = & N \beta \tilde{a}^{\rm IG}(\rho,T) + \beta A^{\rm res}(\rho,T) \label{eq:finalnew}
\end{eqnarray}
where $\tilde{a}^{\rm IG}_{\rm intra}(T)$ is the intramolecular contribution to the Helmholtz energy of a single molecule, calculated from
\begin{equation}
\beta \tilde{a}^{\rm IG}_{\rm intra}(T) = -\ln \left\{\int \exp[-\beta u_{\rm intra}({\bf s})]{\bf ds}\right\}
\label{eq:IGintra}
\end{equation}

Eqs. (\ref{eq:Aexact}) and (\ref{eq:finalnew}) are identical, since they are statements of the definition of a residual property.  The fundamental difference lies in the details of the way in which $\beta A^{\rm res}(\rho,T)$ is constructed in the rigid and flexible FF cases.   We show the consequences of this fact for general thermodynamic properties in the next section, which differ for volumetric and caloric properties.

\subsection{Volumetric vs. Caloric Property Calculations for Flexible FFs} \label{sec:2.3}
The system pressure, $P$, is obtained from Eq. (\ref{eq:finalnew0}) as
\begin{equation}
\beta P  =  -\left(\frac{\partial \beta A}{\partial V}\right)_{T,N} = \frac{N}{V} - \left(\frac{\partial \beta A^{\rm total}_{\rm config}}{\partial V}\right)_{T,N} \label{eq:pressure}
\end{equation}
Comparing Eqs. (\ref{eq:pressure0}) and (\ref{eq:pressure}), we see that $\beta P$ and other volumetric properties obtained from it can be calculated {\it in the case of a flexible FF in the same way as for a rigid FF by simply replacing $u_{\rm inter}({\bf r}_{ij})$ in Eq. (\ref{eq:rigidQ}) by the total FF $u_{\rm total}({\bf z_i})$ of Eq. (\ref{eq:2})}.

On the other hand, caloric properties entail a different treatment, due to the appearance of the term $N \beta \tilde{a}^{\rm IG}_{\rm intra}(T)$ in Eq. (\ref{eq:finalnew0}).  For example, for the internal energy $U$, Eq. (\ref{eq:finalnew0}) gives
\begin{eqnarray}
U \equiv \left(\frac{\partial \beta A}{\partial \beta}\right)_{V,N} & = & N \tilde{u}^{\rm IG}(T) + U^{\rm total}_{\rm config}(\rho,T) - N \tilde{u}_{\rm intra}^{\rm IG}(T)   \label{eq:Ungerer0}\\
 & = & N \tilde{u}^{\rm IG}(T) + U^{\rm inter}_{\rm config}(\rho,T) + U^{\rm intra}_{\rm intra}(\rho,T)  - N \tilde{u}_{\rm intra}^{\rm IG}(T) \\
& = &  U^{\rm IG}(T) + U^{\rm res}
\end{eqnarray}
where $\tilde{u}^{\rm IG}(T)$ is the separately obtained ideal gas value of the internal energy per molecule, and $\tilde{u}_{\rm intra}^{\rm IG}$ is the intramolecular contribution to the internal energy of a single molecule.

The molar IG, residual and total internal energy values are obtained from a simulation by
\begin{eqnarray}
u({\rm sim}) & = & u^{\rm IG}(T) + \left(\frac{N_{\rm Av}}{N}\right)U^{\rm inter}_{\rm config}  + \left(\frac{N_{\rm Av}}{N}\right)U^{\rm intra}_{\rm config} - N_{\rm Av}\tilde{u}^{\rm IG}_{\rm intra}(T)   \label{eq:Ungerer}\\
& = &  u^{\rm IG}(T) + u^{\rm inter}_{\rm config} + u^{\rm intra}_{\rm config} - u^{\rm IG}_{\rm intra}(T) \label{eq:Ucontributions1}\\
& = &  u^{\rm IG}(T) + u^{\rm res}({\rm sim}) \label{eq:Ucontributions2}
\end{eqnarray}
where $N_{\rm Av}$ is Avogadro's number. The total and residual molar enthalpy values are
\begin{eqnarray}
h({\rm sim}) & = & h^{\rm IG}(T) + \left(\frac{N_{\rm Av}}{N}\right)U^{\rm inter}_{\rm config}  + \left(\frac{N_{\rm Av}}{N}\right)U^{\rm intra}_{\rm config} - N_{\rm Av}\tilde{u}^{\rm IG}_{\rm intra}(T) \nonumber \\
& & + P\left(\frac{N_{\rm Av}}{N}\right)\langle V \rangle_{\rm total} -RT \label{eq:Ungererh}\\
& = & h^{\rm IG}(T) + u^{\rm inter}_{\rm config} + u^{\rm intra}_{\rm config} - u^{\rm IG}_{\rm intra}(T) + Pv - RT \label{eq:Ungererh1}\\
& = & h^{\rm IG}(T) + h^{\rm res}({\rm sim}) \label{eq:Ungererh2}
\end{eqnarray}
where $v$ is the molar volume.

In terms of simulation quantities (with FF units of kJ mol$^{-1}$),
\begin{eqnarray}
u^{\rm inter}_{\rm config} & = & \langle{\cal{U}}_{\rm inter}({\bf z})\rangle_{\rm total} \\
u^{\rm intra}_{\rm config} & = &  \langle{\cal{U}}_{\rm intra}({\bf z})\rangle_{\rm total} \\
u^{\rm intra}_{\rm config} & = &  \langle{\cal{U}}_{\rm total}({\bf z})\rangle_{\rm total} \\
u_{\rm intra}^{\rm IG}(T) & = & \langle u_{\rm intra}({\bf s})\rangle_{\rm IG}
\end{eqnarray}
where $\langle\ldots \rangle_{\rm total}$ denotes the configurational average over states of the system occurring with probability proportional to $\exp(-{\cal U}_{\rm total}/(RT)$), and $\langle\ldots\rangle_{\rm IG}$ denotes the configurational average for the simulation of a single molecule over states of the system occurring with probability proportional to $\exp(-\beta {\cal U}_{\rm intra}/(RT))$.

Although experimental values of $u$ and $h$ require the specification of a reference state, the heat capacities may be directly measured experimentally.  These are given from Eqs. (\ref{eq:Ungerer})-(\ref{eq:Ungererh2}) by
\begin{eqnarray}
c_V({\rm sim}) & = &  c_V^{\rm IG}(T) + \left(\frac{\partial u^{\rm inter}_{\rm config}}{\partial T}\right)_{v} + \left(\frac{\partial u^{\rm intra}_{\rm config}}{\partial T}\right)_{v}
-\left(\frac{d u^{\rm IG}_{\rm intra}(T)}{dT}\right) \label{eq:cv}\\
  & = &  c_V^{\rm IG}(T) + \left(\frac{\partial u^{\rm inter}_{\rm config}}{\partial T}\right)_{v} + \left(\frac{\partial u^{\rm intra}_{\rm config}}{\partial T}\right)_{v} -c_V^{\rm IG,\;intra}(T) \\
& = & c_V^{\rm IG}(T) + c_V^{\rm res}({\rm sim})
\end{eqnarray}
\begin{eqnarray}
c_P({\rm sim}) & = & c_P^{\rm IG}(T) + \left(\frac{\partial u^{\rm inter}_{\rm config}}{\partial T}\right)_{P} + \left(\frac{\partial u^{\rm intra}_{\rm config}}{\partial T}\right)_{P} + P\left(\frac{\partial v}{\partial T}\right)_{P} -c_V^{\rm IG,\;intra}(T) -R  \label{eq:Ungerercp} \\
& = & c_P^{\rm IG}(T) + c_P^{\rm res}({\rm sim}) \label{eq:Ungerercp1}
\end{eqnarray}
The ideal gas quantities $c_V^{\rm IG}(T)$ and $c_P^{\rm IG}(T)$ are obtained separately using thermochemical tables or quantum mechanical calculations, and the ideal-gas heat capacity $c_V^{\rm IG,\;intra}(T)$ due to the intramolecular part of the FF can be calculated from the simulation of a single molecule.

Previous workers~\cite{Jorgensen1986,Lagache2001,Lagache2004,Ungerer2005,Ungerer2007,Biscay2011,Caleman2012,Ungerer2015} have focussed on $h({\rm sim})$ and $c_P({\rm sim})$, and all have generally omitted the terms in Eqs. (\ref{eq:Ungererh1}) and (\ref{eq:Ungerercp}) involving the volume on the basis that they are very small in the liquid phase; we will examine its numerical contribution later when discussing our simulation results.  The only groups to correctly and explicitly include the terms involving $u^{\rm IG}_{\rm intra}$ in $h({\rm sim})$ and in $c_P({\rm sim})$ are those of Escobedo and Chen~\cite{Escobedo2001} and of Ungerer \ea \, \cite{Ungerer2015}, but their treatments were very brief and did not consider the general case presented here.

Several groups~\cite{Raabe2010a,Jorgensen1986,Lagache2001,Lagache2004,Ungerer2005,Ungerer2007,Biscay2011} have used only the terms involving $u^{\rm inter}_{\rm config}$ in Eqs. (\ref{eq:Ungerercp}) and (\ref{eq:Ungerercp1}), equivalent to the approximations
\begin{eqnarray}
u^{\rm intra}_{\rm config} & = & u^{\rm IG}_{\rm intra}(T)  \label{eq:IGapprox} \\
\left(\frac{\partial u^{\rm intra}_{\rm config}}{\partial T}\right)_{P} &  = & c_V^{\rm IG,\;intra}(T) \label{eq:IGapprox2}
\end{eqnarray}
in the above expressions, which we refer to as approximation A$_1$.  We refer to the approximation that omits terms involving $u^{\rm IG}_{\rm intra}$ (thus performing the calculations as for a rigid force field whose total potential is the sum of the intermolecular and intramolecular contributions) as approximation A$_0$.
$c_P^{\rm res}({\rm sim})$ and its aforementioned approximations are:
\begin{eqnarray}
c^{\rm res}_P({\rm sim}) & = & \left(\frac{\partial u^{\rm intra}_{\rm config}}{\partial T}\right)_{P} + \left(\frac{\partial u^{\rm inter}_{\rm config}}{\partial T}\right)_{P} + P\left( \frac{\partial v}{\partial T}\right)_P - c_V^{\rm IG,\;intra}(T) -R \label{eq:CpCorrect} \\
c^{\rm res}_P(A_0) & = & \left(\frac{\partial u^{\rm inter}_{\rm config}}{\partial T}\right)_{P} + \left(\frac{\partial u^{\rm intra}_{\rm config}}{\partial T}\right)_{P} + P\left( \frac{\partial v}{\partial T}\right)_P -R \label{eq:CpA0}\\
c^{\rm res}_P(A_1) & = & \left(\frac{\partial u^{\rm inter}_{\rm config}}{\partial T}\right)_{P}  + P\left( \frac{\partial v}{\partial T}\right)_P  -R \label{eq:CpA1}
\end{eqnarray}

Another approximation to the total $c_P$ value is based on the fluctuation of the system Hamiltonian in an $(NPT)$ simulation~\cite{Allen1987}:
\begin{eqnarray}
c_P({\rm classical})  & = & \frac{d}{dT}\left(u^{\rm inter}_{\rm config} + u^{\rm intra}_{\rm config} + Pv + E_k\right) \label{eq:Caleman0}\\
& = & \frac{1}{RT^2}\sigma^2(u^{\rm inter}_{\rm config} + u^{\rm intra}_{\rm config} + Pv + E_k)
\label{eq:Caleman1}
\end{eqnarray}
where $\tilde{v}$ is the volume per particle, $E_k$ is the kinetic energy per particle and $\sigma^2(x)$ is the variance of $x$ within the simulation.  Caleman \ea~\cite{Caleman2012} have performed benchmark MD simulation results for several properties of a large number of organic liquids, including the total $c_P$, for which they used this approximation.  On the reasonable assumption that $E_k$ in an MD simulation is uncorrelated with the remaining quantities, and setting $E_k=f/2RT$, where $f$ is the number of degrees of freedom of the FF, $c_P({\rm classical})$ can be expressed as
\begin{equation}
c_P({\rm classical})  = \frac{d}{dT}\left(u^{\rm inter}_{\rm config} + u^{\rm intra}_{\rm config} + Pv \right) + \left(\frac{f}{2}\right)R
\label{eq:Calemanx}
\end{equation}
The derivatives in Eqs. (\ref{eq:CpCorrect})-(\ref{eq:CpA1}) can be obtained in an $NPT$ simulation by numerical differentiation of the indicated quantities, or also by means of fluctuation quantities~\cite{Hill1986}.  The expressions for $c_P^{\rm res}({\rm sim})$ and its approximations are summarized below in terms of fluctuation quantities.
\begin{eqnarray}
c^{\rm res}_P({\rm sim}) & = & \frac{1}{RT^2}\sigma^2(u^{\rm intra}_{\rm config} + u^{\rm inter}_{\rm config}+ Pv) - c_V^{\rm IG,\;intra}(T) -R  \label{eq:CpCorrect1}\\
c^{\rm res}_P(A_0) & = & \frac{1}{RT^2}\sigma^2(u^{\rm intra}_{\rm config} + u^{\rm inter}_{\rm config}+ Pv) - R\\
c_P^{\rm res}(A_1) & = & \frac{1}{RT^2}
\left[
\sigma^2\left(u^{\rm inter}_{\rm config}\right) +
{\rm cov}(u^{\rm inter}_{\rm config},u^{\rm intra}_{\rm config}) +
{\rm cov}(u^{\rm inter}_{\rm config},Pv)
\right]
 -R
\label{eq:Jorgensen2}\\
c_P^{\rm res}({\rm classical}) & =  & \frac{1}{RT^2}\sigma^2(u^{\rm inter}_{\rm config} + u^{\rm intra}_{\rm config}+ Pv
) +\left(\frac{f}{2}\right)R  -c_P^{\rm IG}(T)\label{eq:Caleman3}
\end{eqnarray}
where ${\rm cov}(x,y)$ is the covariance of $x$ and $y$ within the simulation ensemble, and $c_V^{\rm IG,\;intra}(T)$ is given by Eq. (\ref{eq:IGapprox2}) and by the fluctuation expressions
\begin{equation}
c_V^{\rm IG,\;intra}(T) = \frac{1}{RT^2}\left [\langle u^2_{\rm intra}({\bf s})\rangle_{\rm IG} - \langle u_{\rm intra}({\bf s})\rangle_{\rm IG}^2\right] \equiv \frac{1}{RT^2}\sigma^2(u^{\rm IG}_{\rm intra}({\bf s})) \label{eq:CpIGx}
\end{equation}

Finally, we briefly consider properties arising from a difference across phases of a property at a given value of $T$.  An example is the enthalpy change, $\Delta_{12}h$, given by
\begin{eqnarray}
\Delta_{12}h & = & h^{(2)} - h^{(1)}\\
 & = & [u^{(2)}_{\rm intra} + u^{(2)}_{\rm inter}] - [u^{(1)}_{\rm intra}+ u^{(1)}_{\rm inter}] +  P(v^{(2)} - v^{(1)}) \label{eq:Hvap}
\end{eqnarray}
Using approximation A$_0$ for each phase yields the correct simulation result for $\Delta_{12} h$, since the $u^{\rm IG}_{\rm intra}$ terms in each phase mutually cancel.    However, using only intermolecular contributions (as in approximation A$_1$) for each phase assumes that $u_{\rm intra}$ is the same for each phase.  This is probably not unreasonable if each of the phases in question is solid or liquid.

We note in passing that Eqs. (\ref{eq:finalnew0}) and (\ref{eq:finalnew}) provide a working definition of both a volumetric property and a caloric property, in the context of a flexible FF.  A caloric property requires the use of the term $N \beta \tilde{a}^{\rm IG}$ in Eq. (\ref{eq:finalnew0}), whereas a volumetric property does not.

\section{Systems Studied, Molecular Models and Simulation Details}

We consider liquid state calculations both for the relatively simple molecule 2-propanol: ${\rm CH_3CH(OH)CH_3}$, which was originally studied by Jorgensen~\cite{Jorgensen1986}, and for the more complex molecule monoethanolamine MEA: HO(CH$_2$)$_2$NH$_2$, an important solvent in carbon capture processes~\cite{Tan2016c}. 2-propanol was modelled by the United-Atom OPLS FF of Jorgensen and MEA was modelled by the All-Atom OPLS FF of Caleman \ea~\cite{Caleman2012}.  All bond lengths were fixed in both cases and all bond bending angles were also held fixed for 2-propanol.  The FF parameters are given in the cited papers.

We calculated $h^{\rm res}$  and  $c_P^{\rm res}$ over a range of temperatures at $P=1.01325$ bar for 2-propanol and $P=1$ bar for MEA.  For the experimental values, we used the correlations given below; the values in parentheses indicate the $T$ ranges quoted in the indicated references.  We remark that the value of the 2-propanol $c_P$ correlation of Katayama~\cite{Katayama1962} at 298.15 K of 162.7 J mol$^{-1}$ K$^{-1}$ agrees well with the more recent single value at 298.15 K of 161.2 J mol$^{-1}$ K$^{-1}$ of Roux \ea~\cite{Roux1980}).
\begin{eqnarray}
c_P^{\rm IG}({\rm 2-propanol})& = & 25.535 + 0.21203T + 5.3492 \times 10^{-5}T^2 - 1.4727 \times 10^{-7}T^3 \nonumber \\
& & + 4.9406 \times 10^{-11}T^4;\;\;(100,1500) \label{eq:YawsIGprop}\\
c_P({\rm 2-propanol}) & = & 35.542 + 1.735\times 10^{-2}(T-273.15) +6.941\times 10^{-4}(T-273.15)^2; \nonumber \\
& & (273.15,333.15)\\
c_P^{\rm IG}({\rm MEA}) & = & -0.555 +0.37003 T -3.1976 \times 10^{-4}T^2 + 1.5834\times 10^{-7}T^3 \nonumber \\
& &  - 3.2344\times 10^{-11}T^4;\;\;(298,1500) \label{eq:YawsIGMEA}\\
c_P({\rm MEA}) & = & 79.86 + 0.289T;\;\;(303.15,393.15)
\end{eqnarray}
$c_P^{\rm IG}({\rm 2-propanol})$ and $c_P^{\rm IG}({\rm MEA})$ are from Yaws~\cite{Yaws1999}, $c_P({\rm 2-propanol})$ is from Katayama~\cite{Katayama1962}, and $c_P({\rm MEA})$ is from Rayer \ea~\cite{Rayer2012c}.

For 2-propanol, $NPT$ Monte Carlo simulations were implemented using the Cassandra~1.1 suite~\cite{Shah2011} with the OPLS FF of Jorgensen~\cite{Jorgensen1986} over the temperature range (268.15 K, 338.15 K). For all simulations, $P=1.01325$~bar and $N=500$. Dispersion corrections were applied for the Lennard-Jones interactions and Ewald summation was used to treat the long-range electrostatic forces, with both cutoffs set to 9\,\AA.  The number of MC steps (trial move attempts) in production runs was 131,072,000 with the exception of $T=298.15$\,K, where 524,288,000 configurations were generated.   After each 1000 steps, the energy and volume were stored to calculate the averages of the quantities contributing to the approximations for $h^{\rm res}$, and their simulation uncertainties were calculated using the block average method~\cite{Allen1987,Frenkel2002}.  The simulations were run on a 8 Xeon X3565 @ 3.0 GHz CPU, taking approximately 50 hours per state point.

For MEA, NPT molecular dynamics simulations at $P = 1$  bar and $N = 500$ were implemented using GROMACS 5.1.3~\cite{GROMACS2015} using the OPLS FF of Caleman et al. \cite{Caleman2012} over the temperature range (288.15K, 433.15K) using the leap-frog integration algorithm and a time step of 2 fs.  All bond lengths were kept fixed at their equilibrium values consistent with the force field parameters using the LINCS algorithm~\cite{Hess1997,Hess2008}. Following an initial configuration generated by random insertion, we employed energy minimization using the conjugate gradient algorithm to avoid molecular overlaps. An equilibration stage includes a 2ns run using a v-scaling thermostat and a Berendsen barostat~\cite{Berendsen1984}, followed by a 2ns run using a Nos\'{e}-Hoover~\cite{Nose1984,Hoover1985} thermostat and a Rahman-Parinello barostat~\cite{Parrinello1981}. The production stage consists of a 20ns using a Nos\'{e}-Hoover thermostat and a Rahman-Parinello barostat. The time constant $\tau_t = 1$ ps was used for all thermostats and $\tau_p = 5$ ps for all barostats.  The compressibility for the barostats was $5\times 10^{-5}$ bar$^{-1}$. Dispersion corrections were applied for the LJ interactions and particle-mesh Ewald summation was used to treat the long-range electrostatic forces. The cutoff radii for the LJ and Columbic potentials were 1.1nm and the cutoff radius for the neighbour list was 1.1 nm; the latter was updated every 10 time steps. Since GROMACS treats some intra-molecular terms as inter-molecular quantities, we wrote our own code to calculate $u_{\rm inter}^{\rm config}$ and $u_{\rm intra}^{\rm config}$ from the configurations generated by GROMACS.  All GROMACS simulations were run on a GPU cluster (NVIDIA Tesla K20m GPU + 16 Xeon E5-2680 v2 @ 2.8 GHz CPU), taking approximately 4 hours per state point.

$c_V^{\rm IG, intra}(T)$ for 2-propanol was calculated using the MATLAB~\cite{MATLAB} script shown in Fig. \ref{fig:MATLAB}. It was calculated for MEA by simulating a single molecule in Cassandra 1.2, since we found this approach to be more more precise than using GROMACS, due to the latter's inherent temperature fluctuations.  The ideal gas MEA molecule was fixed at the centre of a cubic box of volume $V = 1$ nm$^3$. $10^6$ MC steps were used for a production run, with each run takes less than 10 minutes on a desktop (2 GHz Intel Core i7 CPU).

At each state point, we calculated all quantities and their uncertainties from 10 independent simulation runs.

\section{Results and Discussion}

\subsection{2-propanol}

The simulation results for the quantities contributing to $u^{\rm res}({\rm sim})$ in Eqs. (\ref{eq:Ucontributions1}) and (\ref{eq:Ucontributions2}) for 2-propanol at $P=1.01325$ bar are given in Table \ref{Udatapropanol}.  As indicated in the table caption, the contribution of the $Pv$ term to $h^{\rm res}({\rm sim})$ in Eqs. (\ref{eq:Ungererh1}) and (\ref{eq:Ungererh2}) is very small and can be neglected.  $u^{\rm inter}_{\rm config}$ is seen to be the dominant contribution to $u^{\rm res}({\rm sim})$ and to its approximation $A_0$ (which omits $u^{\rm IG}_{\rm intra}$), and is the sole contribution to its approximation $A_1$.  $u^{\rm intra}_{\rm config}$ and $u^{\rm IG,\;intra}_{\rm config}$ are shown as open symbols in Fig. \ref{fig:UintraBoth}.

\begin{figure} %3
 \centering
\includegraphics[scale=1.0]{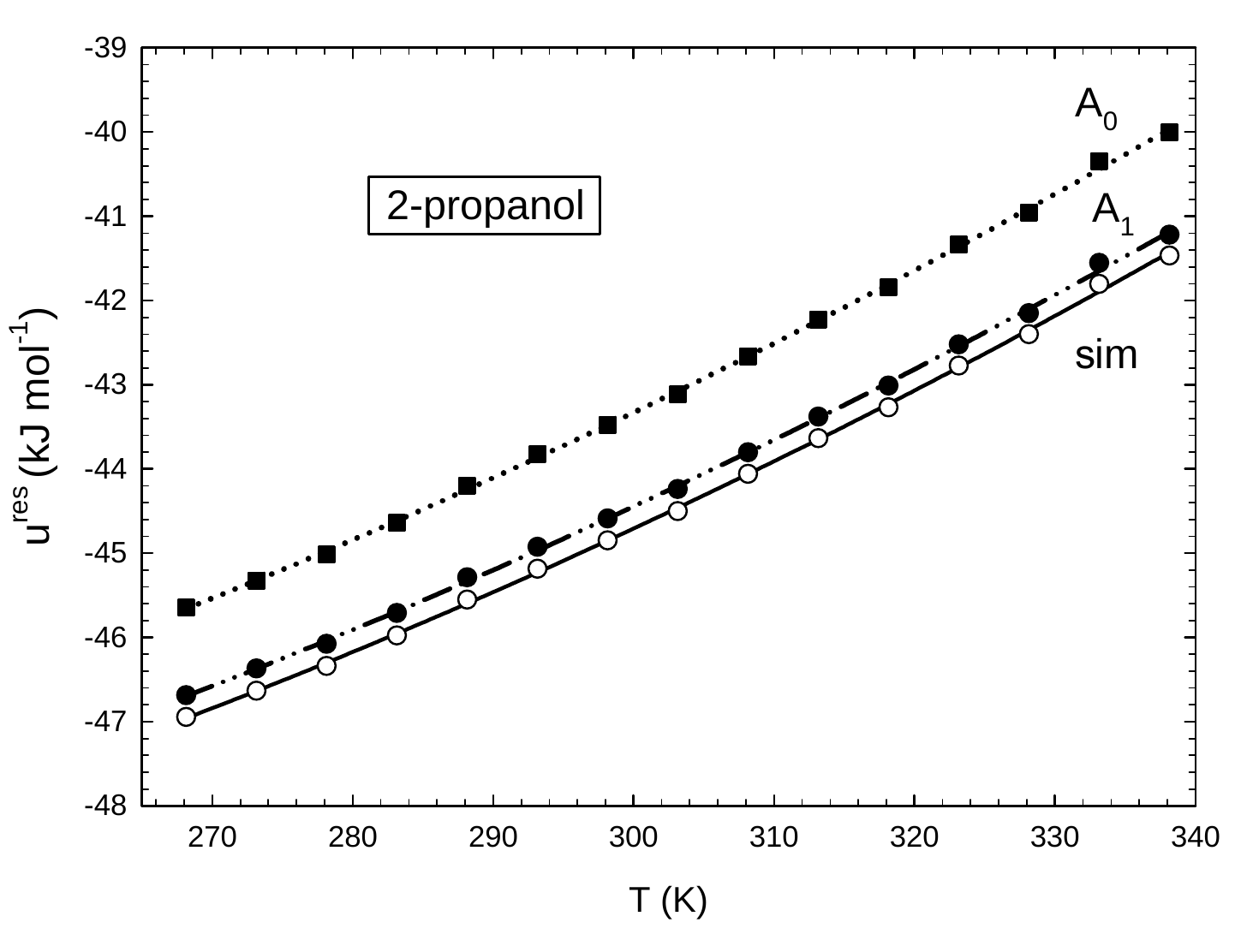}
 \caption{The 2-propanol molar residual internal energy, $u^{\rm res}({\rm sim})$, in  Eqs. (\ref{eq:Ucontributions1}) and (\ref{eq:Ucontributions2}) and its approximations, calculated using the force field of Jorgensen~\cite{Jorgensen1986} at $P=101.325$ kPa.  The simulation values are indicated by open circles. Approximation $A_0$ (filled squares) omits the $u^{\rm IG}_{\rm intra}(T)$ term, and approximation $A_1$ (filled circles) sets $u^{\rm intra}_{\rm config}= u^{\rm IG}_{\rm intra}(T)$.  The curves are drawn using the regression coefficients in Table \ref{Udatapropanol2} and smoothed exact results for $u^{\rm IG}_{\rm intra}(T)$ in Table \ref{Udatapropanol}. The simulation standard deviations lie within the symbol sizes.}
 \label{fig:UresProp}
\end{figure}

The temperature dependence of $u^{\rm res}({\rm sim})$ and its approximations is shown in Fig. \ref{fig:UresProp}.  Due to the near coincidence of the values of $u^{\rm intra}_{\rm config}$ and $u^{\rm IG,\;intra}_{\rm config}$, $A_1$ is seen to be an excellent approximation to $u^{\rm res}({\rm sim})$.  In contrast, $A_0$, which performs the calculation of $u^{\rm res}$ as if 2-propanol were a rigid molecule and thereby omits the quantity $u^{\rm IG}_{\rm intra}$ in Eq (\ref{eq:Ucontributions1}), is a relatively poor approximation.

As noted in Section \ref{sec:2.3}, $c_P^{\rm res}({\rm sim})$ and its approximations can be calculated either by numerical differentiation wrt $T$ of $h^{\rm res}$, or by the use of fluctuation formulae. Proceeding by the first route, we fitted the values of $u^{\rm total}_{\rm config}$ and $u^{\rm inter}_{\rm config}$ to quadratic functions
\begin{equation}
y = a + bT + cT^2 \label{eq:hfit}
\end{equation}
and calculated $c_P^{\rm res}({\rm sim})$ and its approximations analytically from the relevant derivatives; the standard deviation of the prediction at each $T_i$ may be calculated from the relevant regression parameters and covariance data given in Table \ref{Udatapropanol2}, via
\begin{equation}
\sigma^2 (y^{\prime}) = \sigma^2(b) + 4T_i^2 \sigma^2(c) + 4T_i{\rm cov}(b,c) \label{eq:sigmafit}
\end{equation}

$c_V^{\rm IG,\;intra}(T)$ for a single molecule of 2-propanol, shown in the upper pane of Fig. \ref{fig:Cvintra}, was calculated essentially exactly using the MATLAB script of Fig. \ref{fig:MATLAB}, and also from a regression of the form  of  Eq. (\ref{eq:hfit}) for verification of the appropriateness regression approach used for $u^{\rm total}_{\rm config}$ and $u^{\rm inter}_{\rm config}$.  We found that the $c_V^{\rm IG,\;intra}(T)$ results obtained via regression differed from the exact values by less than 0.04 J mol$^{-1}$ at all temperatures.

\begin{figure} %5
 \centering
\includegraphics[scale=1.0]{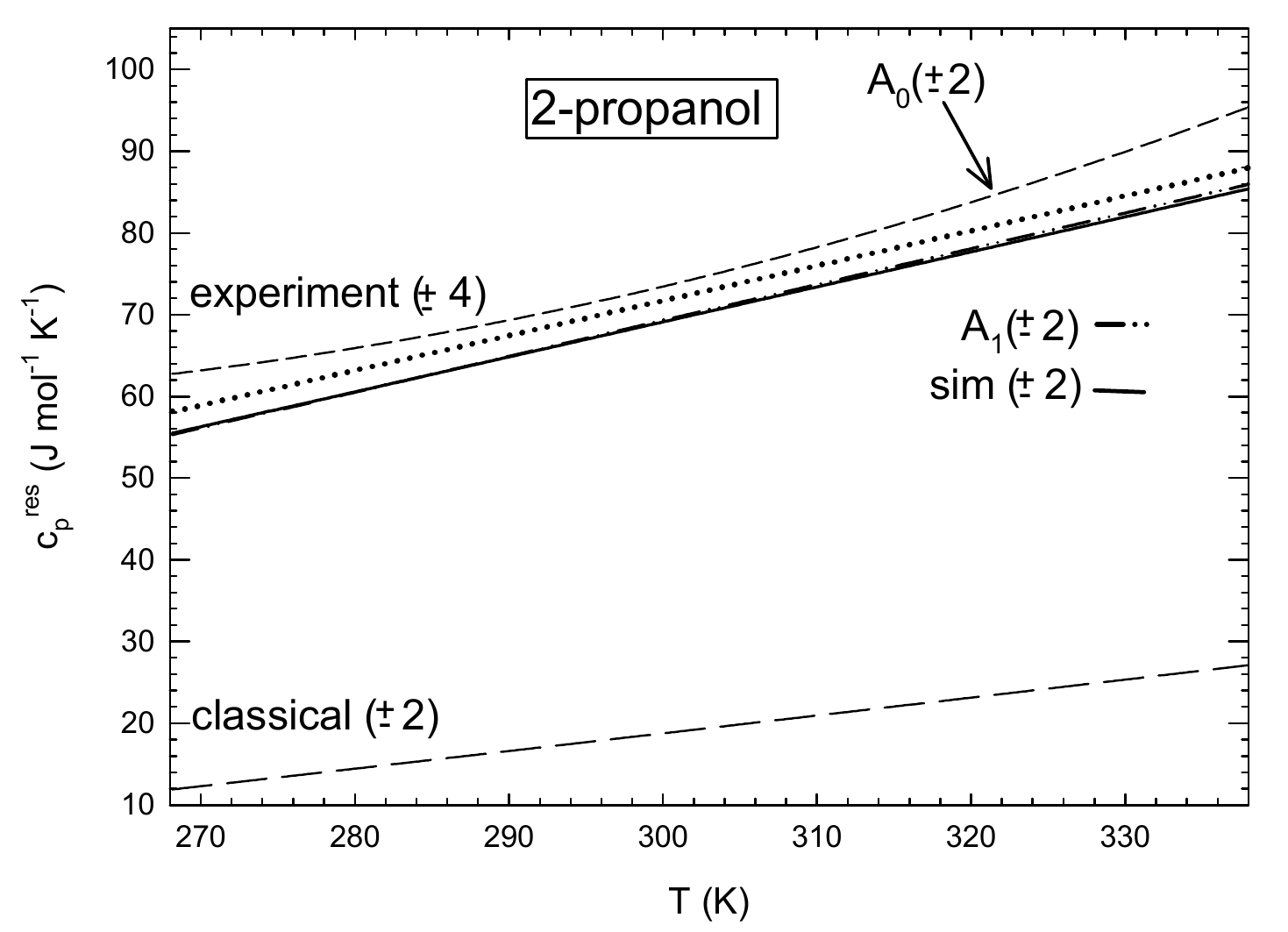}
 \caption{The residual heat capacity, $c_P^{\rm res}({\rm sim})$, of 2-propanol using the OPLS force field of Jorgensen~\cite{Jorgensen1986} at $P=1$ atm., as given by Eq. (\ref{eq:Ungerercp}) and its approximations $c_P^{\rm res}(A_0)$, $c_P^{\rm res}(A_1)$ and $c_P^{\rm res}({\rm classical})$ described in the text. The curves are calculated from the regression coefficients in Table \ref{Udatapropanol2} and the equations described in the text.  The experimental curve shown for comparison is calculated from the total $c_P$ values of Rayer \ea~\cite{Rayer2012c} and $c_P^{\rm IG}$ values from the correlation of Yaws~\cite{Yaws1999}.  The numbers in parentheses denote the standard deviations of the indicated quantities, calculated from the covariance matrix of the parameters in the regression in Table \ref{Udatapropanol2} as described in the text, and for the experimental results it is a subjective estimate based on the scatter of the literature results for the experimental $c_P$ values.}
 \label{fig:Cpres}
\end{figure}

Fig. \ref{fig:Cpres} shows the temperature dependence of $c_P^{\rm res}$ for 2-propanol and its approximations from Eqs. (\ref{eq:CpCorrect})-(\ref{eq:CpA1}) and from Eq. (\ref{eq:Calemanx}) in conjunction with the $c_P^{\rm IG}(T)$ data of Yaws~\cite{Yaws1999}, using the regression coefficients in Table \ref{Udatapropanol2}, and $f=7$ for the Jorgensen FF.  The agreement of the $A_1$ results with $c_P^{\rm res}({\rm sim})$ is excellent, and even those of $A_0$ are is in reasonable agreement with $c_P^{\rm res}({\rm sim})$. This arises from the relatively small difference between $u^{\rm intra}_{\rm config}$ and $u^{\rm IG}_{\rm intra}$ in Fig. \ref{fig:UintraBoth}.  The $c_P^{\rm res}({\rm classical})$ results are seen to be in poor agreement with those of $c_P^{\rm res}({\rm sim})$ and the other approximations; this is likely due to the internal rotations of methyl groups, which ar absent in a united atom FF.  We finally remark that although $c_P^{\rm res}({\rm sim}), A_0$ and $ A_1$ are all in reasonable agreement with the experimental results, only comparison of $c_P^{\rm res}({\rm sim})$ and the experimental curve is a relevant indication of the quality of the FF in its ability to predict this quantity.

We remark in passing that an ``internally consistent" use of the classical approximation includes its application to the ideal-gas case, which yields
\begin{equation}
c_P^{\rm IG}({\rm classical}) = \left(\frac{d u^{\rm IG}_{\rm intra}}{dT}\right) + \left(1+ \frac{f}{2}\right)R \label{eq:Caleman4}
\end{equation}
Subtracting this from the total value of $c_P({\rm classical})$ of Eq. (\ref{eq:Calemanx}) yields a result that matches $c_P^{\rm res}({\rm sim})$  of Eq. (\ref{eq:CpCorrect}).

Finally, we can make contact in passing with the earlier 2-propanol calculations of Jorgensen~\cite{Jorgensen1986} at 298.15 K and 1.01325 bar.  His value (energy units are in cal) of $u^{\rm IG}_{\rm intra}=0.328$ (no uncertainty is given) is in agreement with our converted Table \ref{Udatapropanol} value of 0.329, his value of $u^{\rm intra}_{\rm config}=0.276\pm 0.002$ is in agreement with our converted value of $0.266\pm 0.01$, and his value of $u^{\rm inter}_{\rm config}=-10.61\pm0.03$ is very close to our converted value of $-10.66\pm 0.01$.  Finally, Jorgensen's result for $c_P^{\rm res}(A_1)$ is equivalent to the following approximation to $c_P^{\rm res}(A_1)$, which neglects the covariance terms in Eq. (\ref{eq:Jorgensen2}):
\begin{equation}
c_P^{\rm res,\,Jorgensen} (A_1)= \frac{1}{RT^2}\sigma^2\left(u^{\rm inter}_{\rm config}\right) - R \label{eq:Jorgensen1}
\end{equation}
He obtained the value $13.39 \pm 1.8$ cal mol$^{-1}$ K$^{-1}$, compared with our $c_P^{\rm res}(A_1)$ value of $16.33 \pm 0.14$ cal mol$^{-1}$ K$^{-1}$.  (Using Jorgensen's $c_P^{\rm IG}(298.15)$ value of 21.21 cal mol$^{-1}$ K$^{-1}$, our value for the total $c_P$ for the liquid from Eq. (\ref{eq:Jorgensen2}) is $37.54 \pm 0.14$, cal mol$^{-1}$ K$^{-1}$ vs his value of $34.6 \pm 1.8$ cal mol$^{-1}$ K$^{-1}$.)

\subsection{MEA}

\begin{figure} %6
 \centering
\includegraphics[scale=1.0]{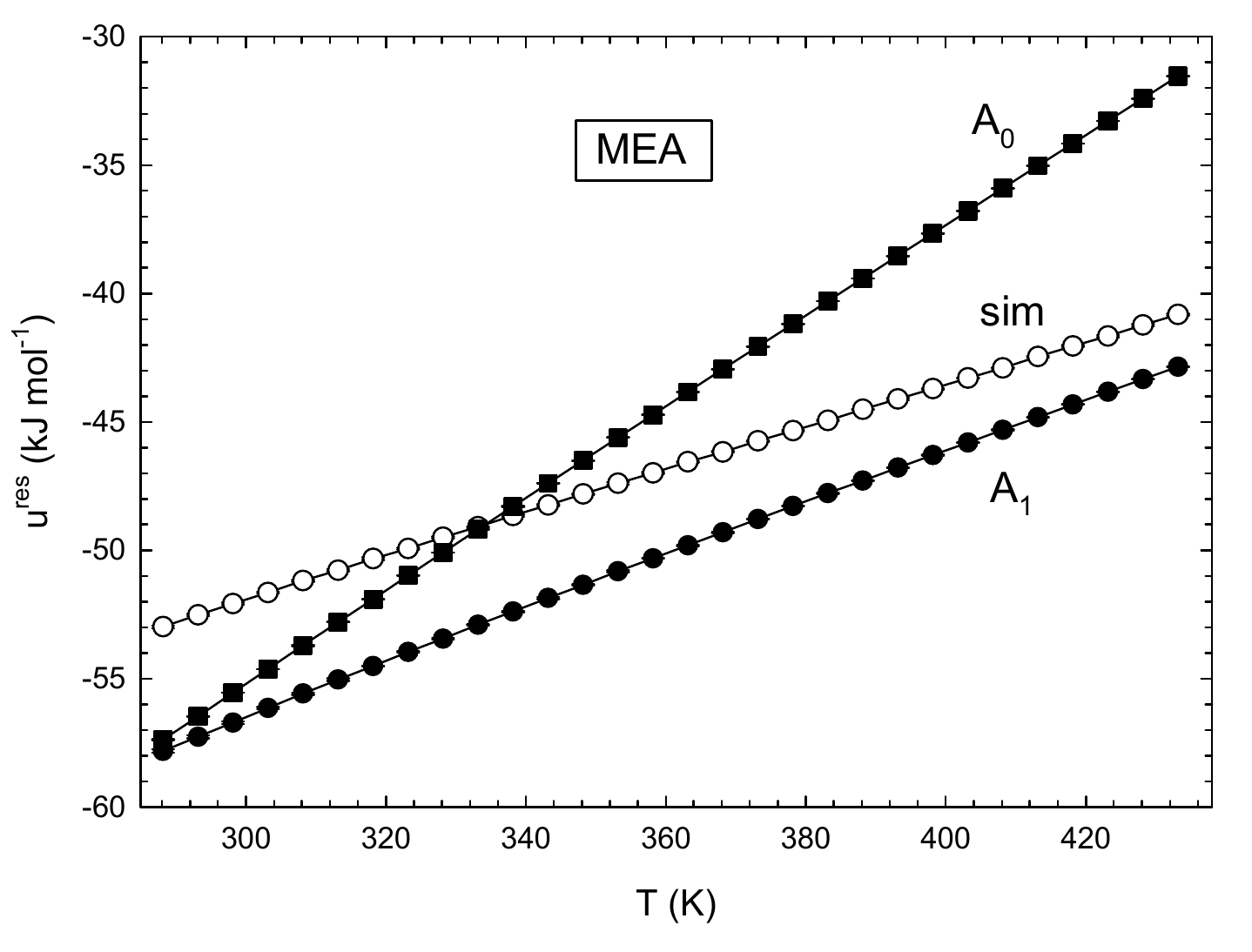}
 \caption{The MEA molar residual internal energy, $u^{\rm res}({\rm sim})$, in  Eqs. (\ref{eq:Ucontributions1}) and (\ref{eq:Ucontributions2}) and its approximations, calculated using the force field of Caleman \ea~\cite{Caleman2012} at $P=1$ bar.  $u^{\rm res}({\rm sim})$ (open circles) is given by $u^{\rm total}_{\rm config}- u^{\rm IG}_{\rm intra}(T)$. Approximation $A_0$ (filled squares) omits the $u^{\rm IG}_{\rm intra}(T)$ term, and approximation $A_1$ (filled circles) sets $u^{\rm intra}_{\rm config}= u^{\rm IG}_{\rm intra}(T)$. The simulation standard deviations lie within the symbol sizes.}
 \label{fig:UresMEA}
\end{figure}

The simulation results for the quantities contributing to $u^{\rm res}({\rm sim})$ in Eqs. (\ref{eq:Ucontributions1}) and (\ref{eq:Ucontributions2}) for MEA at $P=1$ bar are given in Table \ref{UdataMEA1}, and the temperature dependence of $u^{\rm res}({\rm sim})$ and its approximations are shown in Fig. \ref{fig:UresMEA}.   As in the case of 2-propanol, $u^{\rm inter}_{\rm config}$ is the dominant contribution to $u^{\rm res}({\rm sim})$ and its approximations, but for MEA the $u^{\rm intra}_{\rm config}$ and $u^{\rm IG}_{\rm intra}$ data show a greater temperature dependence and mutually differ more markedly than for the simpler 2-propanol molecule.  Also, for MEA the latter quantity is always greater than the former, opposite to the case for 2-propanol.  Both behaviours are shown by the filled symbols in Fig. \ref{fig:UintraBoth}.  $A_1$ is generally a reasonable approximation to the correct results, becoming increasingly more accurate with increasing temperature.  The agreement of approximation $A_0$ with $u^{\rm res}({\rm sim})$ is excellent at the lowest temperatures, but diverges increasingly at higher temperature values.

The MEA simulation results for $c_P^{\rm res}({\rm sim})$ and its approximations from Eqs. (\ref{eq:CpCorrect})-(\ref{eq:CpA1}) and from Eq. (\ref{eq:Calemanx}) in conjunction with the $c_P^{\rm IG}(T)$ data of Yaws~\cite{Yaws1999} are given in Table \ref{UdataMEA2}. (For the MEA FF used, $f= 3\times 11-10=23$).  The lower panel of Fig. \ref{fig:Cvintra} shows the temperature dependence of $c_V^{\rm IG,\;intra}$, which is qualitatively very different over the temperature ranges considered from the 2-propanol results shown in the upper panel.  Its magnitude is also significantly larger than is the case for 2-propanol.

\begin{figure} %7
 \centering
\includegraphics[scale=1.0]{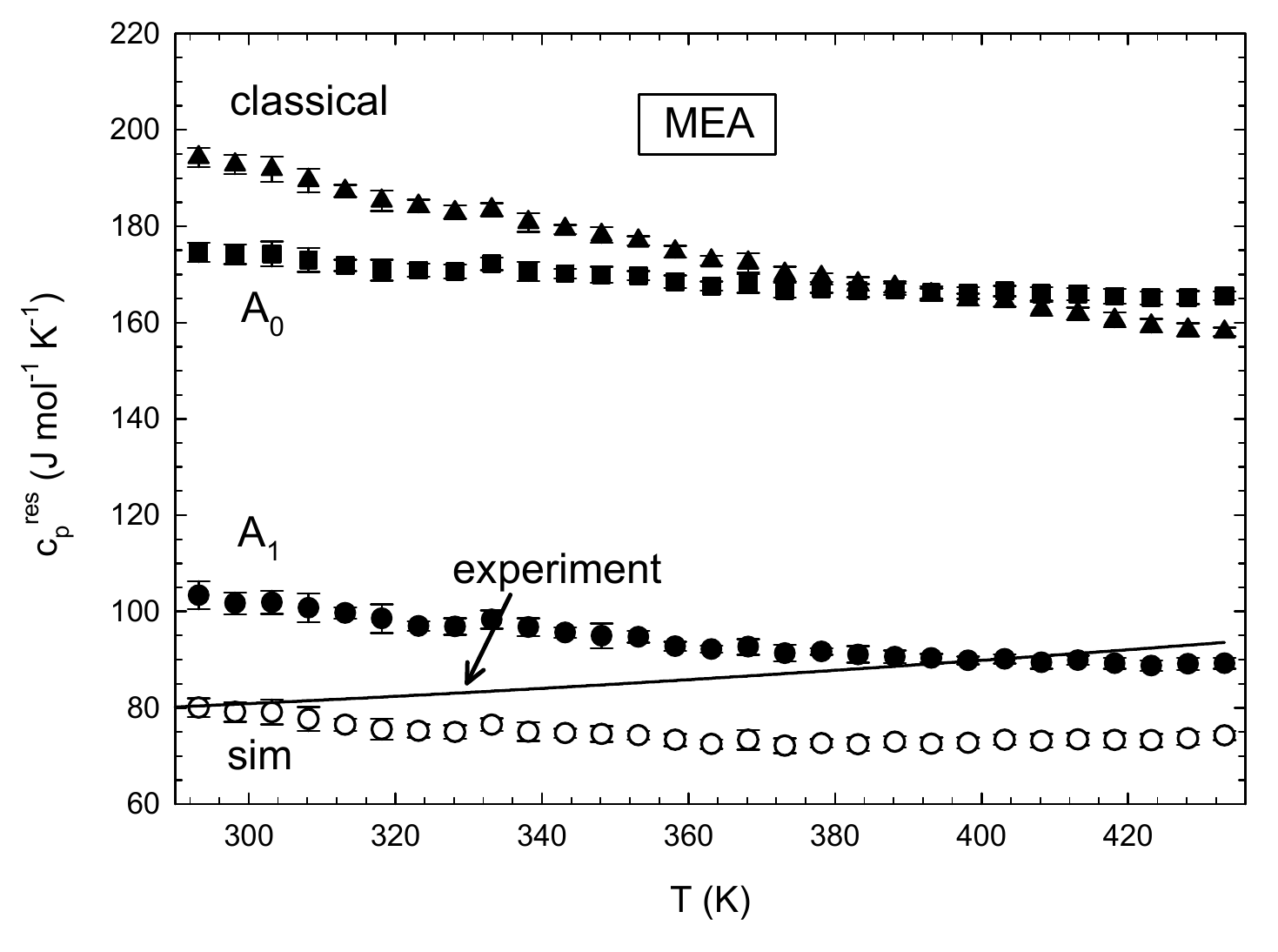}
 \caption{The residual heat capacity, $c_P^{\rm res}({\rm sim})$, of MEA using the OPLS force field of Caleman \ea~\cite{Caleman2012} at $P=1.01325$ bar, as given by Eq. (\ref{eq:Ungerercp}), and its approximations $c_P^{\rm res}(A_0)$, $c_P^{\rm res}(A_1)$ and $c_P^{\rm res}({\rm classical})$ using the data of Table \ref{UdataMEA2}.  All results neglect the contribution of the $Pv$ term, which is very small and within the simulation uncertainties.  The experimental curve shown for comparison is calculated from the total $c_P$ values of Katayama~\cite{Katayama1962} and $c_P^{\rm IG}$ values from the correlation of Yaws~\cite{Yaws1999}.}
 \label{fig:CpresMEA}
\end{figure}

The MEA simulation results for $c_P^{\rm res}({\rm sim})$ and its approximations are compared with each other and with the experimental values in Fig. \ref{fig:CpresMEA}. As indicated in the figure caption, the contribution of the $Pv$ term is very small and can be neglected,  similarly to the case of 2-propanol. As for 2-propanol, approximation $A_1$ is superior to $A_0$, and both are superior to $A_0$, which is not too different from the $c_P^{\rm res}({\rm classical})$ result.  The agreement of $c_P^{\rm res}({\rm sim})$ with the experimental curve is excellent at the lowest temperatures, but diverges below it at higher temperatures, where $A_1$ is better.

\section{Conclusions}

\begin{enumerate}
\item We have derived the correct procedures to calculate, by means of molecular simulation using a classical force field (FF), the total value (incorporating ideal-gas (IG) contributions obtained separately using thermochemical tables or from quantum mechanical calculations) of a thermodynamic property of a fluid whose molecules are modelled by a FF expressed as a sum of intramolecular and intermolecular terms.
\item The total value of a volumetric property is calculated by adding the separately obtained IG value to a residual value calculated using the same methodology as for a rigid force field, by substituting the rigid intermolecular FF in the algorithm by the total FF including the intramolecular and intermolecular terms.  The calculation of the total value of a caloric property requires the additional calculation and incorporation of the simulation value of the relevant property for a single molecule containing only the intramolecular component of the FF.
\item
    An interphase caloric property resulting from the difference of a property's value between two phases at the same $T$, is correctly calculated using approximation $A_0$ for each phase. Approximation $A_1$, which calculates the property in each phase using only the intermolecular FF contributions, is equivalent to assuming that the intramolecular contributions to the property in each phase are identical.  This is likely a reasonable approximation if the phases are solid or liquid.
\item We have illustrated the approach for the calculation of the residual internal energy, residual enthalpy, and the residual heat capacity at constant pressure, $c_P^{\rm res}$, for the case of the relatively simple 2-propanol molecule, at $P=1.01325$ bar from $T=268.15$ K to $T=338.15$ K, and for the more complex monoethanolamine molecule, an important solvent used in CO$_2$ capture, at $P=1$ bar from $T=288.15$ K to $T=433.15$ K.  We have compared the correct results with those of three approximations, with emphasis on $c_P$.
\item Total values of a caloric property should be calculated by means of the methodology described in this paper.  However, on the basis of our $c_P$ calculations, approximation $A_1$ yields not unreasonable results, and is superior to both $A_0$ and to the classical result.
\begin{enumerate}
\item For the state points considered, approximation $A_0$ for $c_P$, which performs the calculation by replacing the intermolecular FF by the total FF in the rigid FF algorithm, differs from the value of $c_P^{\rm res}({\rm sim})$ for 2-propanol in the range (2.3\%, 5.1\%), and for MEA in the range (103\%, 116\%).
\item Approximation $A_1$ for $c_P$, which performs the calculation assuming that the intramolecular contribution to the residual internal energy in the liquid state is unchanged from its value in the ideal gas state, differs from the value of $c_P^{\rm res}({\rm sim})$ for 2-propanol in the range (0.3\%, 0.7\%), and for MEA in the range (9\%, 31\%).
\item $c_P^{\rm res}$ calculated from the ``classical" approximation to the total $c_P$, based on the fluctuation of the total Hamiltonian in an MD simulation~\cite{Allen1987}, from which literature ideal-gas values $c_P^{\rm IG}(T)$ are subtracted, differ from those of $c_P^{\rm res}({\rm sim})$ for 2-propanol in the range (68\%, 78\%) and for MEA in the range (94\%, 141\%).
\end{enumerate}
\end{enumerate}

\section{Acknowledgments}

We thank Mr. Qihao Liu for performing the MATLAB calculations.  Support for this work was provided by the Natural Sciences and Engineering Research Council of Canada (Strategic Program Grant STPGP 479466), the SHARCNET (Shared Hierarchical Academic Research Computing Network) HPC consortium (http://www.sharcnet.ca) and the Czech Science Foundation (Grant No. 15-19542S).

\clearpage

%\bibliography{FlexibleFFs}

\clearpage

\clearpage
\begin{table} %1
    \centering
    \caption{Results as a function of temperature for the contributions to the 2-propanol molar residual internal energy, $u^{\rm res}$, at $P=1.01325$ bar in Eqs. (\ref{eq:Ucontributions1}) and (\ref{eq:Ucontributions2}) in Eqs. (\ref{eq:Ungererh1}) and (\ref{eq:Ungererh2}), and values of $\langle u_{\rm intra}({\bf s})^2\rangle_{\rm IG}$ used in the calculation of $c_V^{\rm IG,\;intra}$ in Eq. (\ref{eq:CpIGx}).  $u^{\rm IG}_{\rm intra}$  and $\langle u_{\rm intra}({\bf s})^2\rangle_{\rm IG}$ were calculated by the MATLAB Script of Fig. \ref{fig:MATLAB}, and can be considered to be exact to the number of digits given.  The $u^{\rm inter}_{\rm config}$ and $u^{\rm intra}_{\rm config}$ values are simulation results, whose standard deviations obtained by the block average method~\cite{Allen1987,Frenkel2002}, are approximately 0.05.  Thee value of the $Pv$ term in Eq. (\ref{eq:Ungererh1}) is less than 0.01 kJ mol$^{-1}$.
    }
\bigskip
    \begin{tabular}{ccccc}
    \hline\\[-.3cm]
$T$	&	$u^{\rm inter}_{\rm config}$	&	$u^{\rm intra}_{\rm config}$	&	$u^{\rm IG}_{\rm intra}$ & $\langle u_{\rm intra}({\bf s})^2\rangle_{\rm IG}$\\[.1cm]
(K)   &   (kJ mol$^{-1}$)                   &  (kJ mol$^{-1}$)                    &    (kJ mol$^{-1}$)  & (kJ$^2$ mol$^{-2}$)      \\
\hline\\[-.3cm]
268.15	&	-46.689	&	1.039	&	1.298 & 3.373\\
273.15	&	-46.370	&	1.048	&	1.311 & 3.428\\
278.15	&	-46.079	&	1.063	&	1.325 & 3.481\\
283.15	&	-45.712	&	1.072	&	1.338 & 3.533\\
288.15	&	-45.289	&	1.088	&	1.351 & 3.584\\
293.15	&	-44.927	&	1.104	&	1.364 & 3.634\\
298.15	&	-44.589	&	1.114	&	1.376 & 3.683\\
303.15	&	-44.241	&	1.126	&	1.388 & 3.731\\
308.15	&	-43.804	&	1.144	&	1.400 & 3.778\\
313.15	&	-43.379	&	1.155	&	1.411 & 3.824\\
318.15	&	-43.009	&	1.162	&	1.422 & 3.868\\
323.15	&	-42.523	&	1.183	&	1.433 & 3.912\\
328.15	&	-42.152	&	1.197	&	1.444 & 3.955\\
333.15	&	-41.554	&	1.205	&	1.454 & 3.997\\
338.15	&	-41.219	&	1.215	&	1.464 & 4.038\\
\hline
    \end{tabular}
	\label{Udatapropanol}
\end{table}

\clearpage
\begin{table} %2
    \centering
    \caption{2-propanol regression data for unweighted least-squares fits to the indicated quantities as quadratic functions of $T$ of Eq. (\ref{eq:hfit}). The  $u^{\rm inter}_{\rm config}$ and $u^{\rm IG}_{\rm intra}$ values are given Table \ref{Udatapropanol} and $u^{\rm total}_{\rm config}$ is the sum of the values in columns two and three of Table \ref{Udatapropanol}.}
\bigskip
    \begin{tabular}{ccccccccc}
    \hline\\[-.3cm]
Quantity	                  & $a$     &  $\sigma^2(a)$     &	$b$         &	$\sigma^2(b)$ &	$c$         & $\sigma^2(c)$ & ${\rm cov}(b,c)$\\[.1cm]
\hline\\[-.3cm]
$u^{\rm total}_{\rm config}$  & -48.128 & 6.218              & -4.805E-02	& 2.726E-04       & 2.135E-04   & 7.408E-10     & -4.492E-07\\[.1cm]
$u^{\rm inter}_{\rm config}$  & -48.409 & 5.989              & -5.103E-02   & 2.625E-04       & 2.141E-04   & 7.134E-10     & -4.325E-07\\[.1cm]
$u^{\rm IG}_{\rm intra}$      & 0.12452 &   4.094E-05        & 5.963E-03   &  1.795E-09       & -5.922E-06  & 4.877E-15     & 2.957E-12\\[.1cm]
\hline
    \end{tabular}
	\label{Udatapropanol2}
\end{table}

\clearpage
{\setstretch{1.0}
    \begin{center}
\begin{longtable}{cccccccc} %3
    \caption{Simulation results for the contributions to the MEA molar residual internal energy, $u^{\rm res}$, at $P=1$ bar as a function of temperature in Eqs. (\ref{eq:Ucontributions1}) and (\ref{eq:Ucontributions2}) and to the molar residual enthalpy, $h^{\rm res}$, in Eqs. (\ref{eq:Ungererh1}) and (\ref{eq:Ungererh2}).  All results were obtained from 10 independent simulation runs, and the values in parentheses denote one standard deviation.  The value of the $Pv$ term in Eq. (\ref{eq:Ungererh1}) is less than 0.01 kJ mol$^{-1}$.}\\
%\bigskip
%    \begin{tabular}
    \hline
$T$	&	$u^{\rm inter}_{\rm config}$	 & $u^{\rm intra}_{\rm config}$ &	$u^{\rm IG}_{\rm intra}$ \\[.1cm]
(K)   &   (kJ mol$^{-1}$)                   &  (kJ mol$^{-1}$)                    &    (kJ mol$^{-1}$)      \\
\hline
288.15	&	-57.81 (0.07)	&	0.44 (0.06)	&	-4.40 (0.05)	\\
293.15	&	-57.26 (0.06)	&	0.80 (0.06)	&	-3.94 (0.04)	\\
298.15	&	-56.71 (0.05)	&	1.17 (0.05)	&	-3.46 (0.04)	\\
303.15	&	-56.14 (0.04)	&	1.51 (0.04)	&	-2.98 (0.04)	\\
308.15	&	-55.58 (0.04)	&	1.87 (0.04)	&	-2.52 (0.03)	\\
313.15	&	-55.03 (0.05)	&	2.24 (0.05)	&	-2.01 (0.04)	\\
318.15	&	-54.51 (0.02)	&	2.61 (0.02)	&	-1.58 (0.05)	\\
323.15	&	-53.96 (0.02)	&	2.97 (0.02)	&	-1.05 (0.03)	\\
328.15	&	-53.44 (0.03)	&	3.35 (0.02)	&	-0.59 (0.03)	\\
333.15	&	-52.90 (0.03)	&	3.72 (0.03)	&	-0.10 (0.03)	\\
338.15	&	-52.39 (0.02)	&	4.09 (0.03)	&	0.36 (0.04)	\\
343.15	&	-51.86 (0.03)	&	4.47 (0.03)	&	0.85 (0.03)	\\
348.15	&	-51.35 (0.02)	&	4.85 (0.02)	&	1.31 (0.02)	\\
353.15	&	-50.82 (0.03)	&	5.21 (0.03)	&	1.79 (0.05)	\\
358.15	&	-50.31 (0.01)	&	5.59 (0.01)	&	2.27 (0.03)	\\
363.15	&	-49.81 (0.03)	&	5.98 (0.03)	&	2.72 (0.04)	\\
368.15	&	-49.30 (0.02)	&	6.36 (0.02)	&	3.22 (0.03)	\\
373.15	&	-48.79 (0.02)	&	6.73 (0.02)	&	3.68 (0.03)	\\
378.15	&	-48.28 (0.02)	&	7.10 (0.02)	&	4.16 (0.04)	\\
383.15	&	-47.78 (0.02)	&	7.49 (0.02)	&	4.64 (0.04)	\\
388.15	&	-47.29 (0.01)	&	7.88 (0.01)	&	5.10 (0.02)	\\
393.15	&	-46.79 (0.02)	&	8.25 (0.01)	&	5.57 (0.04)	\\
398.15	&	-46.29 (0.02)   &	8.63 (0.01)	&	6.04 (0.03)	\\
403.15	&	-45.80 (0.01)	&	9.02 (0.01)	&	6.51 (0.02)	\\
408.15	&	-45.31 (0.01)	&	9.40 (0.02)	&	6.99 (0.03)	\\
413.15	&	-44.81 (0.01)	&	9.78 (0.01)	&	7.43 (0.03)	\\
418.15	&	-44.32 (0.01)	&	10.16 (0.01)	&	7.88 (0.03)	\\
423.15	&	-43.82 (0.01)	&	10.54 (0.01)	&	8.37 (0.04)	\\
428.15	&	-43.33 (0.01)	&	10.92 (0.01	&	8.81 (0.04)	\\
433.15	&	-42.85 (0.01)	&	11.31 (0.02)	&	9.28 (0.04)	\\
\hline
%    \end{tabular}
	\label{UdataMEA1}
\end{longtable}
\end{center}
}

\clearpage
{\setstretch{1.0}
\begin{center}
\begin{longtable}{ccccccccc} %4
    \caption{
    Simulation results for the MEA molar residual heat capacity at constant pressure, $c_P^{\rm res}$, at $P=1$ bar and its approximations from Eqs. (\ref{eq:CpCorrect1})-(\ref{eq:Caleman3}), and the ideal gas quantity $c_V^{\rm IG,\;intra}$, as a function of temperature.   All heat capacities are in J mol$^{-1}$ K$^{-1}$.  $c_P^{\rm res}({\rm classical})$ was obtained by subtracting the $c_P^{\rm IG}$ value~\cite{Yaws1999} given by Eq. (\ref{eq:YawsIGMEA}) from the GROMACS value of $c_P({\rm classical})$  All results were obtained from 10 independent simulation runs, and the values in parentheses denote one standard deviation.  The contribution of the $Pv$ term in Eq. (\ref{eq:Ungererh1}) is less than 0.001 J mol$^{-1}$ K$^{-1}$ at all temperatures.}\\
    \hline
$T$(K)	&	$c_P^{\rm res}({\rm sim})$	&	$c_P^{\rm res}(A_0)$ 	&	$c_P^{\rm res}(A_1)$    	&	 $c_P^{\rm res}({\rm classical})$  & $c_V^{\rm IG,\;intra}$\\
\hline
288.15	&	81.6 (1.1)	&	176.2 (1.1)	&	106.7 (3.2)	&	197.2 (1.9)	& 94.7 (0.5)\\
293.15	&	80.0 (2.0)	&	174.5 (2.0)	&	103.3 (2.9)	&	193.8 (3.2)	& 94.5 (0.4)\\
298.15	&	79.1 (2.0)	&	174.2 (2.0)	&	101.7 (2.3)	&	191.7 (3.3)	& 95.0 (0.3)\\
303.15	&	79.1 (2.6)	&	174.3 (2.6)	&	101.9 (2.4)	&	191.4 (3.9)	& 95.2 (0.3)\\
308.15	&	77.7 (2.5)	&	173.0 (2.5)	&	100.8 (3.0)	&	189.8 (3.7)	& 95.4 (0.4)\\
313.15	&	76.5 (1.2)	&	171.9 (1.2)	&	99.7 (1.2)	&	186.6 (2.6)	& 95.4 (0.3)\\
318.15	&	75.5 (2.2)	&	170.9 (2.2)	&	98.5 (3.0)	&	184.0 (3.1)	& 95.4 (0.3)\\
323.15	&	75.2 (1.3)	&	170.9 (1.3)	&	97.0 (1.0)	&	183.8 (1.6)	& 95.7 (0.3)\\
328.15	&	75.0 (1.4)	&	170.6 (1.4)	&	96.9 (1.7)	&	182.5 (2.6)	& 95.6 (0.6)\\
333.15	&	76.5 (1.3) 	&	172.2 (1.3)	&	98.3 (1.9)	&	183.3 (2.3)	& 95.8 (0.3)\\
338.15	&	75.0 (2.0)	&	170.6 (2.0)	&	96.7 (1.9)	&	179.9 (3.2)	& 95.6 (0.2)\\
343.15	&	74.7 (1.0)	&	170.2 (1.0)	&	95.6 (1.1)	&	178.4 (2.2) & 95.4 (0.3)	\\
348.15	&	74.6 (1.6)	&	169.9 (1.6)	&	94.9 (2.6)	&	177.9 (3.2)	& 95.4 (0.2)\\
353.15	&	74.3 (0.9)	&	169.8 (0.9)	&	94.7 (1.2)	&	177.4 (1.9)	& 95.4 (0.3)\\
358.15	&	73.3 (1.3)	&	168.5 (1.3)	&	92.7 (0.9)	&	174.5 (2.5) & 95.1 (0.3)	\\
363.15	&	72.5 (1.0)	&	167.6 (1.0)	&	92.2 (0.8)	&	171.6 (2.9)	& 95.1 (0.3)\\
368.15	&	73.3 (2.0)	&	168.1 (2.0)	&	92.7 (1.6)	&	171.9 (2.5)	& 94.8 (0.1)\\
373.15	&	72.1 (1.5)	&	166.7 (1.5)	&	91.3 (1.7)	&	169.1 (3.5) & 94.6 (0.1)	\\
378.15	&	72.6 (0.9)	&	167.0 (0.9)	&	91.7 (0.6)	&	168.1 (1.5)	& 94.4 (0.3)\\
383.15	&	72.4 (1.4)	&	166.6 (1.4)	&	91.1 (1.7)	&	167.8 (1.7)	& 94.2 (0.3)\\
388.15	&	72.9 (1.1)	&	166.9 (1.1)	&	90.6 (1.3)	&	167.6 (2.7) & 94.0 (0.3)	\\
393.15	&	72.5 (1.3)	&	166.3 (1.3)	&	90.3 (0.8)	&	165.7 (3.1)	& 93.7 (0.2)\\
398.15	&	72.7 (1.2)	&	166.1 (1.2)	&	89.8 (0.8)	&	164.6 (2.5)	& 93.4 (0.3)\\
403.15	&	73.4 (1.1)	&	166.6 (1.1)	&	90.2 (0.9)	&	164.0 (2.6)	& 93.2 (0.2)\\
408.15	&	73.1 (1.4)	&	166.0 (1.4)	&	89.4 (1.3)	&	162.9 (2.5)	& 92.9 (0.1)\\
413.15	&	73.4 (1.3)	&	166.0 (1.4)	&	89.9 (1.0)	&	160.8 (2.5)	& 92.5 (0.2)\\
418.15	&	73.3 (1.5)	&	165.5 (1.5) &	89.2 (1.1)	&	159.4 (3.0)	& 92.1 (0.2)\\
423.15	&	73.2 (1.4)	&	165.1 (1.4) &	88.8 (1.1)	&	158.8 (2.8)	& 91.9 (0.3)\\
428.15	&	73.7 (1.4)	&	165.2 (1.4)	&	89.1 (1.3)	&	159.0 (2.2)	& 91.5 (0.3)\\
433.15	&	74.2 (0.9)	&	165.6 (0.9)	&	89.3 (1.1)	&	157.7 (2.0)	& 91.3 (0.2)\\
\hline
	\label{UdataMEA2}
\end{longtable}
\end{center}
}

%\clearpage
\begin{figure} %1
 \centering
\begin{verbatim}
clc
clear
format long
V0 = 0.429;
V1 = 0.784;
V2 = 0.125;
V3 = -0.691;
T = [268.15
273.15
278.15
283.15
288.15
293.15
298.15
303.15
308.15
313.15
318.15
323.15
328.15
333.15
338.15]
R = 8.31446E-3/4.184;
U = @(phi) V0 + 0.5*V1*(1+cos(phi))+0.5*V2*(1-cos(2*phi))+0.5*V3*(1+cos(3*phi));
for i = 1:15
b = @(phi) exp(-1.*U(phi)./(R*T(i)));
Q(i) =  integral(b,0,2*pi);
Uav(i) = 1/Q(i)*integral(@(phi) U(phi).*b(phi),0,2*pi)*4.184;
U_2av(i) = 1/Q(i)*integral(@(phi) U(phi).^2.*b(phi),0,2*pi)*4.184^2;
end
Q = Q'
Uav = Uav'
U_2av = U_2av'
\end{verbatim}

\caption{MATLAB~\cite{MATLAB} script to calculate the quantities $u^{\rm IG}_{\rm intra}\equiv\langle u_{\rm intra}({\bf s})\rangle_{\rm IG}$ and $\langle u_{\rm intra}({\bf s})^2\rangle_{\rm IG}$ in Eq. (\ref{eq:CpIGx}) as functions of temperature for 2-propanol in Eq. (\ref{eq:CpIGx}) using the OPLS force field of Jorgensen~\cite{Jorgensen1986}. }
\label{fig:MATLAB}
\end{figure}

\clearpage
\begin{figure} %2
 \centering
 \includegraphics[scale=1.0]{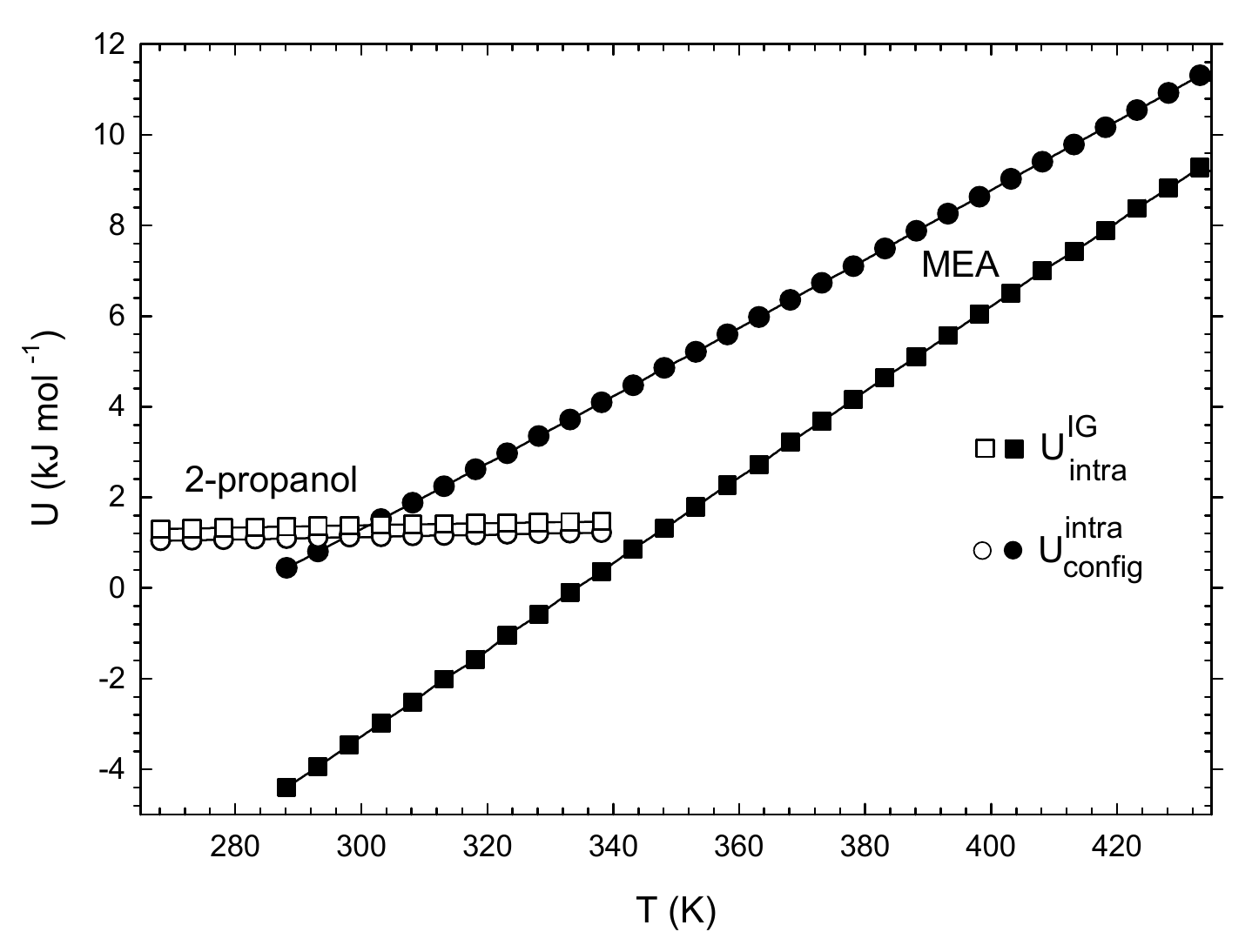}
 \caption{The intramolecular contributions to the molar internal energy, $u({\rm sim})$, in Eq. (\ref{eq:Ucontributions1}).  The data are shown for 2-propanol (open symbols) and for MEA (filled symbols) as a function of temperature.  $u^{\rm intra}_{\rm config}$ for 2-propanol is shown for $P = 1.01325$ bar, and for MEA at $P = 1$ bar, and $u^{\rm IG}_{\rm intra}$ is a function of temperature only.
Apart from $u^{\rm IG}_{\rm intra}$ for 2-propanol, which was calculated using the MATLAB script in Fig. \ref{fig:MATLAB}, the points are simulation results obtained as described in the text.  The curves are drawn as an aid to the eye, and the simulation standard deviations lie within the symbol sizes.}
 \label{fig:UintraBoth}
\end{figure}

\clearpage

\clearpage
\begin{figure} %4
 \centering
\includegraphics[scale=0.8]{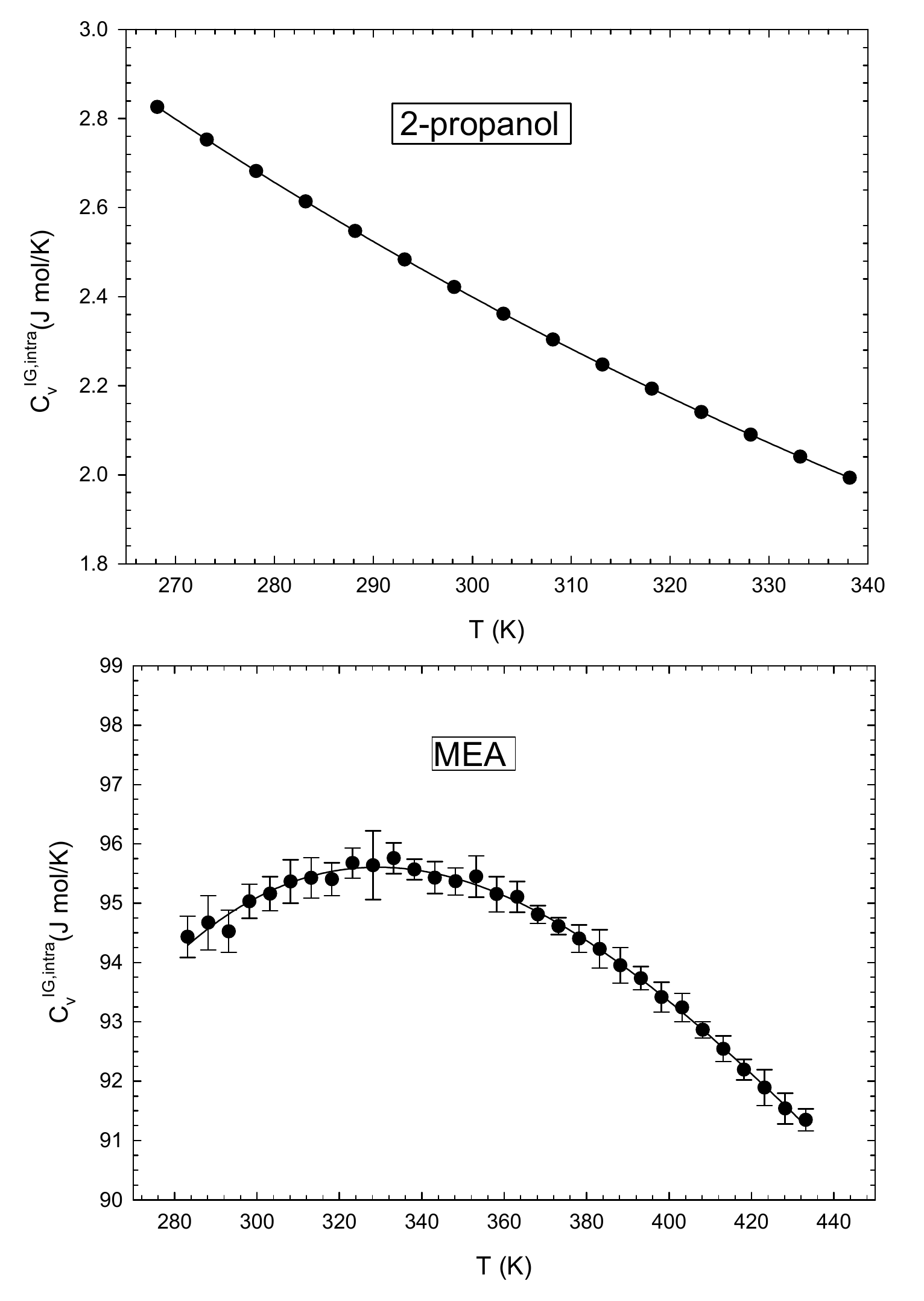}
 \caption{The ideal-gas intramolecular contribution to the molar heat capacity at constant volume, $c_V^{\rm intra,\;IG}(T)$ as a function of temperature.  The upper figure shows data for 2-propanol and the lower figure shows data for MEA.  The 2-propanol data are essentially exact, and were calculated using Eq. (\ref{eq:CpIGx}) and the MATLAB script in Fig. \ref{fig:MATLAB}.  The MEA data were calculated using Cassandra 1.1 as described in the text.  The uncertainties indicate one standard deviation of the results obtained from 10 independent simulation runs.}
 \label{fig:Cvintra}
\end{figure}

\clearpage

\end{document}